# All-Optical, Reconfigurable and Power Independent Neural Activation Function by Means of Phase Modulation

George Sarantoglou. Adonis Bogris, and Charis Mesaritakis

*Abstract*— In this work, we present numerical results concerning an integrated photonic non-linear activation function that relies on a power independent, non-linear phase to amplitude conversion in a passive optical resonator. The underlying mechanism is universal to all optical filters, whereas here, simulations were based on micro-ring resonators (MRRs). Investigation revealed that the photonic neural node can be tuned to support a wide variety of continuous activation functions that are relevant to the neural network architectures, such as the sigmoid and the softplus functions. The proposed photonic node is numerically evaluated in the context of time delayed reservoir computing (TDRC) scheme, targeting the one-step ahead prediction of the Santa Fe series. The proposed phase to amplitude TDRC is benchmarked versus the conventional amplitude based TDRC, showcasing a performance boost by one order of magnitude.

*Index Terms*—photonic neural networks, reservoir computing, photonic integration technologies

## I. INTRODUCTION

Artificial neural networks (ANN) are the main driver behind the accelerating evolution of artificial intelligence, by presenting state of the art performance in a rich variety of tasks such as image recognition, audio and natural language processing [1]. However, the increasing processing power of modern ANNs is accompanied by higher requirements in computational resources, that cannot be met by the development of typical Von – Neumann architectures [2]. In order to circumvent this roadblock, unconventional brain-inspired hardware platforms are investigated, which are liberated from the restrictions of conventional Von – Neumann processors. In this landscape silicon photonics have emerged as a promising candidate for such applications, owing to their wavelength assisted parallelism, inherent linear processing capabilities, marginal power consumption and the alleviation of fan-in/fan-out and bandwidth limitations that plague electronics [3].

Although photonic devices are effective in the case of linear operations [4], [5], [6], they are not yet efficient tools of various on-demand non-linear activation functions, a property that is crucial in the operation of ANNs [7]. All-optical activation functions have been demonstrated in literature using semiconductor optical amplifiers [8] or saturable absorbers [4], non-linear effects triggered by free carrier excitation in micro-ring resonators [9], [10], [11], and semiconductor lasers [12]. In most of these systems, the non-linear response is not tunable or only partially tunable, covering a small number of basic activation functions, whereas the underlying mechanism is bandwidth limited in the case of active materials. In terms of input power, silicon photonic devices present a rather low Kerr coefficient in the order of $10^{-16}$ m$^2$ / W, thus requiring increased input power to be activated [9]. More-over, due to the low optical losses in silicon photonics, non-linear transformations due to free carrier dispersion and two-photon absorption require large input power (>7 mW [11]). Lower power consumption can be achieved by incorporating phase change materials [13] or Ge-Si structures [14], but such procedures require non-conventional fabrication methods. On the other hand, III-V materials despite their higher Kerr coefficient (e.g., $3 - 5 \times 10^{-15}$ m$^2$ / W for GaAs [15]), are lossy and not CMOS compatible. An alternative route, consists of electro-optical activation functions, where the signal is detected, processed in the electronic domain and re-applied to the optical domain [16], [17]. Despite its simplicity, the electro-optic scheme is hampered in terms of scaling by the modulator technology: characterized either by large physical footprint (lithium niobite - LNOI), low bandwidth (thermal actuators) or increased losses when high speed is required (silicon on insulator modulators) [18]. To make matters worse, the use of such electro-optic schemes in large numbers impose restrictions considering circuit packaging and node control.

In this work, a new method for reconfigurable all-optical activation functions is introduced, based on the non-linear phase to amplitude (PTA) response present in passive optical filters, such as MRRs or ring-loaded Mach Zehnder interferometers (MZIs). The shape of the activation function is governed by the transfer function of the filter, which, in turn, can be tuned by modifying few easy-to-handle hyperparameters, while more importantly PTA conversion is power independent. Similar schemes have been employed in the

This work was supported by the EU Horizon Europe PROMETHEUS Project under Grant 101070195. *(Corresponding author: Charis Mesaritakis).*

George Sarantoglou and Charis Mesaritakis are with the Department of Information and Communication Systems Engineering, University of the Aegean, 83200 Karlovassi Samos, Greece (e-mail: gsarantoglou@aegean.gr; cmesar@aegean.gr).

Adonis Bogris is with the Department of Informatics and Computer Engineering, University of West Attica, 12243 Athens, Greece (e-mail: abogris@uniwa.gr).

past for modulation format conversion (PSK to ASK and vice versa), exploiting the transfer function of an MRR [19]. The organization of the manuscript is as follows: first, the main PTA is introduced followed by a simulation of an MRR, so as to present various different non-linear functions. Afterwards an application scenario is realized based on a TDRC scheme [20], targeting the one-step ahead prediction Sante Fe task [21], which utilizes the PTA non-linearity. The PTA TDRC is compared to a similar TDRC based on conventional amplitude modulation (AM), revealing a boost in performance by one order of magnitude, due to the power free non-linearity. The work concludes with a comparison between the PTA method and other TDRCs in literature, along with a discussion about a future perspective of the PTA in neuromorphic photonics in general.

## II. RECONFIGURABLE ACTIVATION FUNCTION BASED ON PHASE TO AMPLITUDE TRANSITIONS

*A. Phase to Amplitude non-linear conversion*

Since the optical activation function is based on the non-linear PTA mapping, information is encoded at the phase of the optical carrier envelope as:

$$E(t) = \sqrt{P} \; exp \, (j \, 2\pi \, m \, n(t)) \quad (1)$$

Here, $P$ is the input power, $m>0$ is the modulation index and $n(t) \in [-1,1]$ is the normalized signal. Given the central frequency of the carrier $f_c$ and the phase encoded information $\phi(t)$, the frequency of the optical signal is equal to:

$$f(t) = f_c + \frac{1}{2\pi}\frac{d\phi(t)}{dt} = f_c + f_M(t) \quad (2)$$

In general, optical filters are characterized in the spectral domain by a non-linear transfer function $H(f)$, which defines its transmissivity. By changing the frequency of the optical signal by phase modulation, the transmissivity is modified, which amounts to a non-linear PTA response due to the non-linear shape of the transfer function in the spectral domain. A physical interpretation of this process, can be derived by considering the splitting of the optical signal at the input of the optical filter and the interference of the modes after the applications of asymmetrical temporal delays (e.g. unbalanced MZI, MRR). The two modes, due to the phase modulation and the applied temporal delay interfere with a phase difference that varies over time. For this reason, $f_M(t)$ is defined by the gradient of the phase signal, which is determined by the normalized signal, the modulation rate and the modulation depth. This mechanism is independent from input power, meaning that it is highly tailored for low power applications.

The effect of a phase change in the transmissivity of the filter is illustrated for the case of the drop port of an MRR in Fig. 1. The shape of the non-linear response can be reconfigured either by changing the frequency detuning between the optical carrier and the resonance of the filter (Fig.1 – green dot), or by modifying the transfer function of the filter. The first feature can be achieved by configuring a phase shifter placed in the cavity of the filter, whereas the second feature becomes available by the state-of-the-art reprogrammable silicon photonic processors [22].

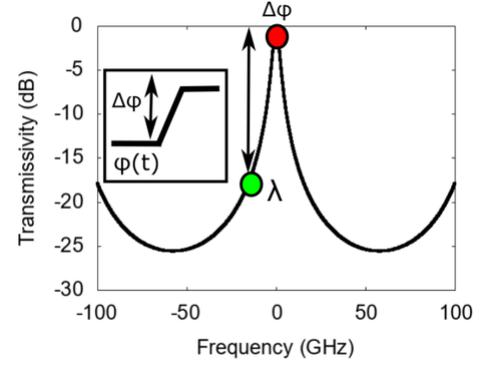

**Fig. 1.** The effect of phase gradients in the transmissivity of an IIR optical filter, based on the drop port of an MRR. The carrier is detuned from the resonance of the filter (green dot). A phase shift $\Delta\phi$ (shown in the inset), causes a change in transmissivity (red dot).

*B. Micro-ring resonator as a reconfigurable photonic neuron*

In order to demonstrate the aforementioned mechanism, an MRR filter is simulated according to the transfer function of the drop port [23]:

$$H_{drop} = \frac{s^2 \sqrt{\zeta}}{c^2 \zeta - 1} \quad (3)$$

Here, $s = \sqrt{k}, c = \sqrt{1-k}$, where k is the coupling coefficient, that is the same for the through and drop ports for the shake of simplicity. Moreover, $\zeta = \gamma \, exp \, (-j \, (2\pi(f + df)T_{ring}))$, where $\gamma$ are the propagation losses, $T_{ring}$ is the round-trip time in the MRR cavity and $df$ is the frequency detuning. The radius of the ring is set at $R=100$ μm, the coupling coefficient is set at $k=0.1$ and the propagation losses are equal to 4 dB/m assuming a Silicon-on-Insulator platform [9]. These choices determine the Q-factor of the cavity. It must be highlighted, that different Q-factors still retain the PTA effect, since the underlying mechanism is universal to optical filters. The time constant of the MRR filter is calculated by the 3dB frequency of the transfer function $f_{3dB}$, as $\tau = 2\pi f_{3db} = 82$ ps.

The input power in (1) is set equal to 1 mW and 100 μW to demonstrate the power independence of the PTA mechanism. In order to evaluate the response of the filter to multiple phase gradients, 40 different signals of the form presented in (1) are inserted in the filter, corresponding to phase transitions $\Delta\phi$ (see inset in Fig.1) ranging from -2πm (i = 1) up to 2πm (i=40). The rise time of the transitions is set to $t_r = 100$ ps, corresponding to a modulation rate of 10 GHz. For each signal, the peak power at the output of the filter is detected. For this scope a photodiode of responsivity 1 A / W is numerically considered and of bandwidth equal to 20 GHz, that is placed at the drop port of the MRR. Both the effects of the shot noise and Johnson noise are simulated. The simulation step is set equal to dt =200 fs.

The shape of the activation function is controlled by two hyperparameters, namely the modulation index m and the frequency detuning df of the MRR filter. The frequency detuning can be set by a phase shifter in the cavity of the MRR, whereas modulation index values up to m=1 can be achieved at 10 Gsa/s by lithium niobite at 20.48 pJ/bit and by indium phosphide modulators at 8 pJ/bit [14]. In Fig.2 three characteristic activation functions f(x) are presented, while the fitting curve is equal to $g(x) = a + bf(((x - \mu))/\sigma)$, where x are the samples and $a, b, \mu, \sigma$ are treated as fitting parameters. The output y is presented in arbitrary units, that result by normalizing the sampled photocurrent values as $y = i/|i|_{max}$, where $|i|_{max}$ is the maximum absolute value recorded among the different signals. The value of $|i|_{max}$ is important, since it describes the extinction ratio of the analog activation function.

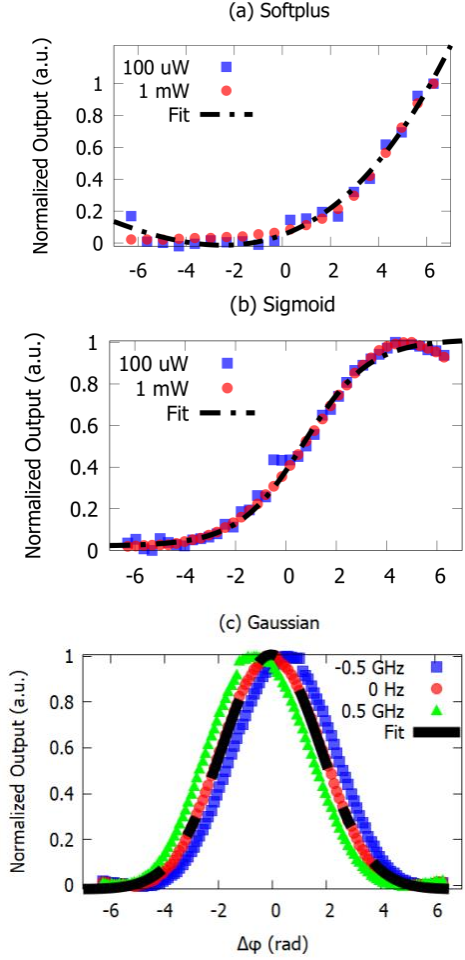

**Fig. 2.** The effect of phase gradients in the transmissivity of an IIR optical filter, based on the drop port of an MRR. The carrier is detuned from the resonance of the filter (green dot). A phase shift Δϕ (shown in the inset), causes a change in transmissivity (red dot).: Different activation function acquired from the drop port of a single MRR node through PTA

In Fig. 2a, the frequency detuning is set equal to df=-10 GHz, whereas the modulation index is set at m=1. The non-linear response in this case resembles a soft-plus activation function of the form $f(x) = log(1 + exp(x))$, which is true for both P=100 μW ($|i|_{max}$ =52.18 μA) and P=1 mW ($|i|_{max}$ = 0.48 mA), confirming that input power does not affect the non-linear process. In Fig. 2b, a sigmoid function of the form $f(x) = [1 + exp(-x)]^{-1}$ is presented, that can be derived by setting m=0.6 and df=-3 GHz. In this case, $|i|_{max}$ = 79 μA for P=100 μW and $|i|_{max}$ = 0.76 mA for P=1 mW. Finally, in Fig. 2c, a Gaussian activation $f(x) = exp(-x^2)$ function is presented that is derived by setting m=1 and df=-0.5,0,0.5 GHz and P=1 mW. Three detuning values are selected to reveal the effect of a perturbation of ±0.5 GHz around 0 Hz, due to external noise effect such as unwanted thermal shifts. For the gaussian analog function, the scaling is equal to $|i|_{max}$ = 1 mA for P=1 mW. As it can be seen, uncertainty with respect to the frequency detuning between the carrier and the resonance do not affect the form of the non-linearity. Consequently, the MRR can recreate multiple activation functions that are independent of the input power and tolerant to phase uncertainty.

### III. TIME DELAYED RESERVOIR COMPUTING BASED ON PHASE TO AMPLITUDE CONVERSION

In order to demonstrate the PTA conversion mechanism in an application-wise scenario, the MRR node is used in a TDRC scheme to provide one-step ahead prediction for the Santa Fe benchmark [20]. The simulated setup is presented in Fig. 3 and it has the same architecture with the TDRC in Ref. [11]. First, the time series $x_{1 \times N_d}$ of length $N_d$ are normalized and multiplied in the digital domain with a mask -matrix $W_{k \times 1}$ that consists of fixed random numbers drawn by the uniform distribution $U(0,1)$. This process performs a dimensionality expansion on the input signal by mapping it to $y_{k \times N_d} = W_{k \times 1} x_{1 \times N_d}$. The i-th column of the output matrix holds k values known as virtual nodes, that will be used from the linear regression, implemented at the front-end, to predict the next point $x_{i+1}$. The matrix $y_{k \times N_d}$ is serialized and it is used by a digital to analog converter (DAC) to implement the electric signal that modulates the optical carrier. The sampling rate SR of the DAC determines the virtual node separation θ. The bandwidth limitation of the modulation is considered by adding a 4th order Butterworth filter, whereas a precision equal to 8 bits is also implemented to include the quantization error.

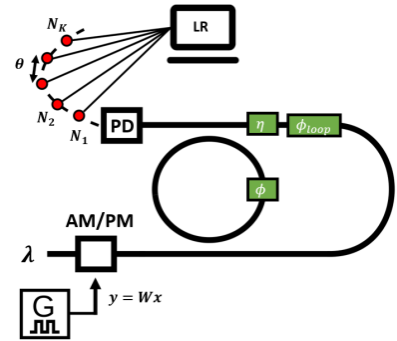

**Fig. 3.** The TDRC setup that is used for the one-step ahead prediction of the Sante Fe chaotic series.

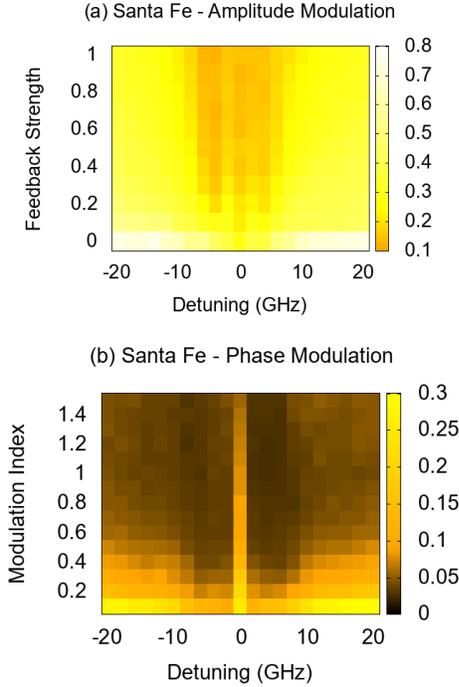

**Fig. 4.** The NMSE as a function (a) of the feedback strength and frequency detuning for AM and (b) of the modulation index and frequency detuning for PM and injection strength equal to 0.5.

Both AM and PM formats are evaluated, with PM presented in (1), whereas AM is given by the following formula:

$$E(t) = \sqrt{P(1 + m\, n(t))} \quad (4)$$

The modulated field is inserted in the photonic RC, which consists of an MRR with an external loop that connects the through with the add port, thus providing the physical memory [10]. The delay of the loop is $T_{loop} = T_{mask}(k+1)/k$, where $T_{mask} = k\theta$. In this way, the loop and the mask are desynchronized, thus imposing interconnectivity between adjacent virtual nodes [24]. The feedback loop can be controlled in terms of its phase $\phi_{loop}$ by including a phase shifter and int terms of its feedback strength $\eta$, by using a variable optical attenuator. The transfer matrix of the MRR is written as [23]:

$$\begin{bmatrix} E_{in} \\ E_{thr} \end{bmatrix} = \frac{1}{-s^2\sqrt{\zeta}} \begin{bmatrix} 1 - c^2\zeta & c(\zeta - 1) \\ c(1 - \zeta) & \zeta - c^2 \end{bmatrix} \begin{bmatrix} E_{drp} \\ E_{add} \end{bmatrix} \quad (5)$$

The transfer function of the TDRC can be written as:

$$H_{TDRC} = \frac{1 - \Phi_{2,2}L}{\Phi_{1,1}(1 - \Phi_{2,2}L) + \Phi_{2,1}\Phi_{1,2}L} \quad (6)$$

Here, $L = \sqrt{\eta}\, exp(j[2\pi f T_{loop} + \phi_{Loop}])$, whereas $\Phi_{ij}$ corresponds to the $(i,j)$ element of the transfer matrix in (5). The photodiode has a bandwidth equal to 0.75×SR, implemented by a 4th order Butterworth filter. The sampling rate of the analog to digital converter (ADC) at the output is set equal to SR, whereas the quantization error at this stage is also included with precision equal to 8-bits. The data are normalized and used by the linear regression in order to perform one-step ahead prediction. The performance of the TDRC is computed in terms of the normalized mean square error (NMSE) [11], defined as:

$$NMSE = \frac{\sum_i (o_i - \hat{o}_i)^2}{N_d \sigma_{\hat{o}}^2} \quad (7)$$

Here, $o_i$ is the $i-th$ estimated output, whereas $\hat{o}_i$ is the target value. The variance of the target values is given by $\sigma_{\hat{o}}^2$. For the following simulations, 2000 samples are used for training and 1000 samples for testing. The parameters k=50 and SR=10 Gsa/s are used (θ=100 ps). The bandwidth of the modulation is set equal to 20 GHz. All subsequent simulations are repeated 10 times for each choice of hyperparameters, so as to compute the mean NMSE along with its standard deviation for different masks $W_{k\times 1}$.

By scanning multiple values for the phase of the feedback loop, its optimum setting is detected as $\phi_{loop}/2\pi = 0.55$ for both AM and PM. The input power is set at 1 mW. For AM, the modulation index is set at m=1 and its performance is evaluated in terms of the frequency detuning $df$ and the feedback strength $\eta$. The results are shown in Fig. 4a. The optimum performance is observed for negative detuning equal to -4 GHz and η=1, returning an NMSE equal to $0.12 \pm 0.019$. The RC achieves a better NMSE compared to the linear shift register, that achieves an NMSE = 0.2. As it is expected, these results are inferior compared to those presented by other TDRC schemes based on Kerr induced non-linearities [11].

In terms of PM, the input is again set at 1 mW. The optimum feedback strength is found at η=0.5. A parametric analysis with respect to the modulation index and frequency detuning is performed, since these terms modify the non-linear PTA conversion (see Fig.2). The results are shown in Fig. 4b. The optimum NMSE = $0.024 \pm 0.004$ is observed for a detuning equal to 4 GHz and a modulation index equal to 1.1. It can be seen that as the modulation index increases, a wide area is observed in terms of frequency detuning, where the system presents NMSE≤ 0.05. The improvement compared to AM can be explained by the non-linear transformation provided by the PTA process. Different results are provided for each choice of the inspected hyperparameters, since they implement different activation functions.

Finally, in Fig. 5 the NMSE as a function of input power is presented both for the PM and the AM schemes. In both cases the performance deteriorates due to the decreased SNR for decreased input power. However, the PM scheme is superior to the AM scheme, both due to its higher SNR and to its non-linear PTA mechanism. Moreover, it is observed that even for -10 dBm, the NMSE of the PM scheme is as low as 0.041 since the non-linear effect is still active, thus demonstrating its power independence.

Compared to an MRR based TDRC with AM and Kerr induced non-linearity [11], the PTA mechanism provides superior results at 7 dB lower input power (NMSE = 0.023 at 0 dBm as opposed to 0.045 at 7dBm) and similar results (NMSE

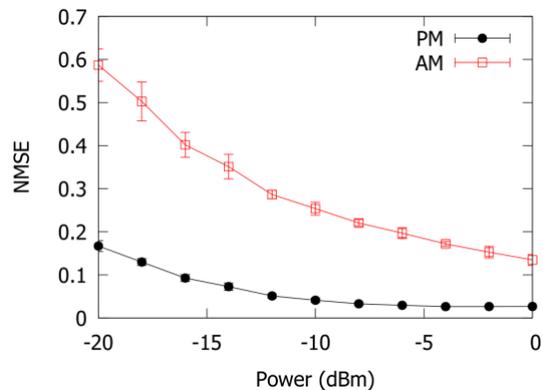

**Fig. 5.** The NMSE as a function of the input Power for AM and PM. The system in each case is set at the optimum bias points derived by the analysis in Fig. 4.

≈ 0.045 at -10 dBm as opposed to 7 dBm), for 17 dB lower input power. Finally, regarding laser based TDRC, although they can provide slightly better results [25], they are bandwidth limited and more importantly not easily integrated with silicon photonics, which is the main platform for neuromorphic photonic systems.

IV. CONCLUSION

A method for replicating any non-linear activation function in the photonic domain has been presented based on PTA conversion in optical filters. Since the underlying mechanism originates from interference in passive optical filters, the non-linearity is power independent. Three different training activation functions are demonstrated with a single MRR. Multiple different activation functions are available simply by choosing different optical filters. The MRR neuron is also used in a TDRC scenario, showcasing its superiority compared to AM modulation schemes in terms of performance and power efficiency. Apart from the RC environment, optical filters utilizing PTA can also be used in convolutional neural networks and feedforward networks, thus providing an additional trainable element, that is the activation function. When deep architectures are considered, PTA non-linearities can used either along with complex ANNs as complex activation functions [25] or along with conventional real-valued ANNs by using all optical ASK to PSK converters [26]. Moreover, due to their power independence, they can provide robustness to the high optical losses that plague state-of-the-art silicon photonic structures [27]. Finally, it has been shown that multi-layer neural networks essentially are polynomial regression models [28]. The degree of the polynomial function is increased at each next layer. The reconfigurability of the PTA mechanism, can be combined with polynomic regression, so as to design single layer photonic networks, to emulate deep structures.

ACKNOWLEDGMENT

This work has received funding from the EU Horizon Europe Program PROMETHEUS under grant agreement 101070195.

**George Sarantoglou** received the Diploma degree in electrical and computer engineering from the University of Patras, Patras, Greece, in 2016. He is currently working toward the Ph.D. degree with the Department of Information and Communication Systems Engineering, University of the Aegean, Karlovassi Samos, Greece. His Ph.D. thesis focuses on the experimental analysis and development of photonic processors for unconventional, bio-inspired information processing, and targeting machine learning applications. His research interests include photonic systems for analog pattern recognition and multisensory applications.

**Adonis Bogris** was born in Athens. He received the B.S. degree in informatics, the M.Sc. degree in telecommunications, and the Ph.D. degree from the National and Kapodistrian University of Athens, Athens, Greece, in 1997, 1999, and 2005, respectively. His doctoral thesis was on all-optical processing by means of fiber-based devices. He is currently a Professor with the Department of Informatics and Computer Engineering, University of West Attica, Aigaleo, Greece. He has authored or coauthored more than 150 articles published in international scientific journals and conference proceedings and he has participated in plethora of EU and national research projects. His current research interests include high-speed all-optical transmissions systems and networks, non-linear effects in optical fibers, all-optical signal processing, neuromorphic photonics, mid-infrared photonics, and cryptography at the physical layer. Dr. Bogris is a reviewer of the journals of the IEEE.

**Charis Mesaritakis** received the B.S. degree in informatics from the Department of Informatics and Telecommunications, National and Kapodistrian University of Athens, Athens, Germany, in 2004, the M.Sc. degree in microelectronics from the Department of Informatics and Telecommunications, Kapodistrian University of Athens, and the Ph.D. degree in the field of quantum dot devices and systems for next generation optical networks from the Photonics Technology and Optical Communication Laboratory, Kapodistrian University of Athens, in 2011. In 2012, he was awarded a European scholarship for post-doctoral studies (Marie Curie FP7-PEOPLE IEF) in the joint research facilities of Alcatel-Thales-Lucent in Paris-France, where he worked on intra-satellite communications. He has actively participated as a Research Engineer/Technical Supervisor in more than ten EU-funded research programs (FP6-FP7-H2020) targeting excellence in the field of photonic neuromorphic computing, cyber-physical security, and photonic integration. He is currently an Associate Professor with the Department of Information and Communication Systems Engineering, University of the Aegean, Mytilene, Greece. He is the author or coauthor of more than 80 papers in highly cited peer reviewed international journals and conferences, two international book chapters. He is a regular reviewer of IEEE.